\documentclass[prl,aps,showpacs,twocolumn,floatfix,superscriptaddress]{revtex4}
\usepackage{eurosym}
\usepackage{amsmath,amsfonts,amssymb,graphics,graphicx,color,bm}

\begin{document}

\title{Composite Boson Mapping for Lattice Boson Systems }
\author{Daniel Huerga}

\author{Jorge Dukelsky}
\affiliation{%
  Instituto de Estructura de la Materia,
  C.S.I.C.,
  Serrano 123, E-28006 Madrid, Spain}%
\author{ Gustavo E. Scuseria }
\affiliation{%
  Department of Chemistry and
  Department of Physics and Astronomy, Rice University, Houston, TX 77005, USA
}%
\begin{abstract}
We present a canonical mapping transforming physical boson operators into
quadratic products of cluster composite bosons that preserves matrix
elements of operators when a physical constraint is enforced.
We map the 2D lattice Bose-Hubbard Hamiltonian into $2\times 2$
composite bosons and solve it within a generalized Hartree-Bogoliubov approximation.
The resulting Mott insulator-superfluid phase diagram
reproduces well Quantum Monte Carlo results.
The Higgs boson behavior in the superfluid phase along the unit density line
is unraveled and in remarkable agreement with experiments.
Results for the properties  of the ground and excited states are competitive
with other state-of-the-art approaches, but at a fraction
of their computational cost. The composite boson mapping here
introduced can be readily applied to frustrated many-body systems
where most methodologies face significant hurdles.
\end{abstract}
\maketitle

{\it Introduction.}-- In the past few years, there has been great
experimental progress on the control and manipulation of cold
atomic gases loaded in optical lattices, leading to quantum simulators
of the Bose-Hubbard model and its Mott insulator to superfluid
transition \cite{Book}. A notable recent experiment has revealed
the Higgs boson behavior across this transition in a 2D optical
lattice \cite{Endres}. There is currently great interest in cold atomic physics
for engineering synthetic gauge fields that induce topological phases
and phase transitions. This can be accomplished using a combination
of laser-induced tunneling with superlattice techniques \cite{Laser},
or by time-periodic shaking of the lattice \cite{Shake}. From the
theoretical perspective, traditional mean-field approaches can describe
the phase diagram of bosonic atoms in lattices of various geometries,
but only qualitatively \cite{Oosten, Huber}. Quantum Monte Carlo (QMC) yields a highly
accurate description of ground state properties at zero and finite
temperatures whenever the system has no frustration \cite{QMC, QMC2}. Static and
dynamic properties have also been studied with the Variational
Cluster Approximation (VCA) \cite{KAL1, KAL2}. Extensions of static mean-field
approaches involving the use of clusters have been considered \cite{Pek, Danshi, Multisite}.
In this work, we introduce a theory that maps cluster subspaces
of the original Fock space onto composite bosons containing the {\it exact}
internal dynamics of the cluster, and whose interactions account
for residual correlations between the clusters. Because the
mapping is canonical, it is then possible to apply standard
many-body techniques to this Composite Boson (CB) Hamiltonian.
In this sense, the method builds upon previous slave-particle theories, extending its realm to clusters along the lines of Hierarchical Mean Field Theory (HMFT)
for quantum magnetism \cite{HMFT}. It could also be considered as an extension to clusters of the on-site slave-boson mapping of Bose-Hubbard model proposed in Ref. \cite {Stoof}. These ideas are here generalized to
interacting bosons systems loaded in optical lattices.
We refer to the resulting method as Composite Boson Mean Field Theory (CBMFT).
We demonstrate that the inclusion of higher order fluctuation terms in the
composite mean-field yields very accurate results.
The CB approach to the Bose-Hubbard model unravels the Higgs boson behavior
along the particle-hole (p-h) symmetry
line and yields remarkable agreement with experimental data \cite{Endres}.

{\it Composite Boson Mapping.}-- Let us start our derivation by decomposing  the original lattice into a perfect tiled cluster lattice (superlattice).
 The cluster states are represented by CBs labeled by a position $R$ in the superlattice and by a set of internal quantum numbers $\alpha$ which constitute a complete and orthonormal basis in the Fock space of the cluster. We propose a quadratic mapping of the boson creation (annihilation) operators $a^{\dagger}_i$ ($a_i$) in terms of these CBs $b^{\dagger}_{R \alpha}$ ($b_{R \alpha}$) as
\begin{equation}
a_i^{\dag}   = \sum\limits_{\alpha \beta } {\left\langle R\alpha  \right|} a_i^\dag  \left| R\beta  \right\rangle b_{R\alpha} ^\dag  b_{R\beta}  ,
\;\;a_i^{}  =\left(a_i^{\dag}\right)^{\dag}, ~~i\in R.
\label{Imag}
\end{equation}

 Let us now
explore the conditions which should be fulfilled by transformation $\left( %
\ref{Imag}\right) $ in order to preserve the canonical bosonic commutation
relations $\left[ a_{i},a_{j}^{\dagger }\right] =\delta _{i,j}$. For $i,j \in R$, we insert the transformation in the commutator and obtain

\begin{eqnarray}
\left[ a_{i},a_{j}^{\dagger }\right] &=&
\sum_{\alpha \beta \beta ^{\prime }}(\left\langle R\alpha \right\vert
a_{i}\left\vert R\beta \right\rangle \left\langle R\beta \right\vert
a_{j}^{\dagger }\left\vert R\beta ^{\prime }\right\rangle \notag\\
&&-\left\langle R\alpha
\right\vert a_{j}^{\dagger }\left\vert R\beta \right\rangle \left\langle
R\beta \right\vert a_{i}\left\vert R\beta ^{\prime }\right\rangle )
b_{R\alpha }^{\dagger }b_{R\beta ^{\prime }}  .
\end{eqnarray}

The satisfaction of the canonical commutation relations relies on
$i)$ resolution of the identity, $\sum_{\beta }\left\vert R\beta
\right\rangle \left\langle R\beta \right\vert =I$, and $ii)$
fulfillment of the physical constraint, $\sum_{\alpha }b_{R\alpha
}^{\dagger }b_{R\alpha }=I$. The latter condition defines the \textit{physical subspace} of the CB Fock space, which has a one to one correspondence with the original Fock cluster space. Alternatively, if $i \in R$ and $j \in R'$ the commutation relation is trivially satisfied due to the commutation of the CBs, $[b_{R \alpha},b^{\dag}_{R'\beta}]=\delta_{RR'} \delta{\alpha \beta}$.

 A direct consequence of the CB mapping is that any operator $\hat{O}_R$ that is an algebraic function of the physical bosons $(a_{i},a^{\dagger}_{j})$
within a single cluster at position $R$ will be mapped to a one-body CB operator, $\hat{O}_R = \sum_{\alpha \beta } {\left\langle R\alpha  \right|} \hat{O}_R \left| R\beta  \right\rangle b_{R\alpha} ^\dag  b_{R\beta}$. This means that the operator $\hat{O}_R$ changes a cluster configuration $\alpha$ into another cluster configuration $\beta$. A formal derivation starting from the mapping $(\ref{Imag})$ in the cluster and using conditions $i)$ and $ii)$ is given in the Supplemental Material (SM).
In the same way, any product of operators belonging to $N$
different clusters will be mapped to an $N$-body operator.
For the sake of simplicity, we will here restrict ourselves to a
density-density interaction that leads to a two-body CB Hamiltonian, since
each density operator is contained in a single cluster,
\begin{equation}
H=\sum_{ij}\left[ t_{ij}a_{i}^{\dagger}a_{j}+V_{ij}n_{i}n_{j}\right]
,~n_{i}=a_{i}^{\dagger}a_{i}.\label{HH}
\end{equation}
 This class of Hamiltonians, with long-range hopping and interactions, covers most of the physical lattice boson models.

We assume a square lattice partitioned into a set of $
M$ clusters, each one at position $R$ of a CB superlattice and containing $L\times L$ sites. Next, we formally map the Hamiltonian using the prescription described above and
rewrite it in terms of CBs labeled by the occupation configuration of
each cluster, $\mathbf{n}\equiv \left\{ n_{1},\dots
,n_{L},\dots n_{L^2}\right\} $,
\begin{eqnarray}
H_{CB} &=&\sum_{R\mathbf{nm}}\left\langle R\mathbf{n}\right\vert
H\left\vert R\mathbf{m}\right\rangle b_{R\mathbf{n}}^{\dagger }b_{R%
\mathbf{m}}  \notag \\
&&+\sum_{ RR^{\prime } }\sum_{\mathbf{nn^{\prime }}%
\mathbf{mm^{\prime }}}\left\langle R\mathbf{n}R^{\prime }\mathbf{n^{\prime }%
}\right\vert H\left\vert R\mathbf{m}R^{\prime }\mathbf{m^{\prime }}%
\right\rangle   \notag \\
&&~~~~~~~~~~~~~~~~~~~~~\times b_{R\mathbf{n}}^{\dagger }b_{R^{\prime }%
\mathbf{n^{\prime }}}^{\dagger }b_{R\mathbf{m}}b_{R^{\prime }\mathbf{%
m^{\prime }}}.
\end{eqnarray}

For reasons that will become clear below, we will perform a
generic unitary
transformation among the CBs, $b_{R\alpha }^{\dagger }=\sum_{\mathbf{n}}U_{R%
\mathbf{n}}^{\alpha }b_{R\mathbf{n}}^{\dagger }$. In this
new basis, the hamiltonian can be written as
\begin{eqnarray}
H_{CB} &=&\sum_{R}\sum_{\alpha \beta }T_{\beta }^{\alpha }(R)~ b_{R\alpha }^{\dagger
}b_{R\beta } \label{Ham}\\
&&+\sum_{ RR^{\prime } }\sum_{\alpha \alpha ^{\prime }\beta
\beta ^{\prime }}W_{\beta \beta ^{\prime }}^{\alpha \alpha ^{\prime
}}(R,R')~ b_{R\alpha }^{\dagger }b_{R^{\prime}{ \alpha^{\prime} }}^{\dagger }b_{R\beta
}b_{R^{\prime}{ \beta^{\prime} }} \notag ,
\end{eqnarray}%
where the intra-cluster $T_{\beta }^{\alpha }(R)$ and the inter-cluster $W_{\beta \beta ^{\prime }}^{\alpha \alpha ^{\prime
}}(R,R')$ matrix elements expressed in the transformed basis encode all the information of the original Hamiltonian. The CB
Hamiltonian $\left( \ref{Ham}\right) $ is an exact image of the original
boson Hamiltonian provided that the physical constraint in each cluster, $\sum_{\alpha }b_{R\alpha }^{\dagger }b_{R\alpha }=I $,
is satisfied. Furthermore, treating
this Hamiltonian by means of standard many-body techniques, we immediately
incorporate quantum correlations inside the cluster in an exact way.

{\it Composite Boson Mean Field Theory.}-- We here treat the CB Hamiltonian
in the Hartree-Bogoliubov approximation. In order
to proceed further, we have to specify the matrix elements of the
initial lattice Hamiltonian. As a first test
of CBMFT, we benchmark the Bose-Hubbard Hamiltonian in a
2D square lattice. Namely, $t_{ij}=-t\delta _{i,j+\mathbf{e}}$
where $\mathbf{e} $ is the unit vector in the lattice directions
$\mathbf{x}$, $\mathbf{y}$, and $V_{ij}=V\delta _{i,j}$ is the
on-site Hubbard repulsion. In what follows, we omit $V$ and measure
all quantities in units of $V$. Assuming a uniform 2D lattice with
translational symmetry, we first perform a Fourier
transform of the CB boson operators $b_{\mathbf{R}\alpha }^{\dagger }=(1/%
\sqrt{M})\sum_{\mathbf{k}}e^{-iL\mathbf{k}\cdot \mathbf{R}}~b_{\mathbf{k}%
\alpha }^{\dagger }$, leading to
\begin{eqnarray}
&&H_{CB} =\sum_{\alpha \beta }T_{\beta }^{\alpha }\sum_{\mathbf{k}}b_{\mathbf{k}%
\alpha }^{\dagger }b_{\mathbf{k}\beta }  \label{FouHam} \\
&&~~~+\frac{1}{M}\sum_{\substack{ \alpha \alpha ^{\prime }\beta \beta ^{\prime }
}}W_{\beta \beta ^{\prime }}^{\alpha \alpha ^{\prime }}\sum_{\substack{
\mathbf{k}_{1}\mathbf{k}_{2}\mathbf{q}}}\gamma _{\mathbf{q}}~b_{\mathbf{k_{1}%
}\alpha }^{\dagger }b_{\mathbf{k_{2}+q}\alpha ^{\prime }}^{\dagger }b_{%
\mathbf{k_{1}+q}\beta }b_{\mathbf{k_{2}}\beta ^{\prime }},  \notag
\end{eqnarray}%
where we have introduced $\gamma _{\mathbf{q}}= \cos (Lq_{x})+\cos
(Lq_{y})$ after a symmetrization of the two-body matrix elements $W_{\beta
\beta ^{\prime }}^{\alpha \alpha ^{\prime }}$ in order to preserve the lattice $C_{4}
$ symmetry. Details on the calculation of these matrix elements
can be found in the SM.
Next, we assume a condensation of the CBs in the $%
\mathbf{k=0},\alpha =0$ state by introducing a shift transformation $b_{%
\mathbf{k=0},\alpha =0}=b_{\mathbf{k=0},\alpha =0}^{\dagger }=\sigma \sqrt{M}
$. This transformation manifestly violates the physical constraint as it induces mixtures with unphysical states. This is a common problem to all slave-particle theories treated in mean-field. However, this mixture is expected to be less severe with increasing cluster sizes, such that in the limit of very large clusters it must be negligible.
Thus, we relax it and impose a global constraint
on the CB density, $\sum_{R}\sum_{\alpha }b_{R\alpha }^{\dagger }b_{R\alpha }=M$.
Transforming to momentum space and shifting, this global physical constraint
becomes
\begin{equation}
\sigma ^{2}+\frac{1}{M}\sum_{\alpha \neq 0}\sum_{\mathbf{k}}b_{\mathbf{k}%
\alpha }^{\dagger }b_{\mathbf{k}\alpha }=1,
\label{Constraint}
\end{equation}
where we have neglected the fluctuations of the condensed $\alpha=0$ CB. Eq. (\ref{Constraint}) defines $\sigma ^{2}$ as the
CB condensate fraction. Inserting the constraint (\ref{Constraint}) by means of a Lagrange multiplier
$\lambda $ in the CB Hamiltonian (\ref{FouHam}) and applying a
mean-field decoupling, we arrive to a quadratic Hamiltonian of the
form
\begin{eqnarray}
H_{MF} &=&H^{(0)}+\sum_{\mathbf{k}} \sum_{\alpha \neq 0, \beta \neq 0}A_{\mathbf{k}\alpha \beta
}b_{\mathbf{k}\alpha }^{\dagger }b_{\mathbf{k}\beta }  \label{MF} \\
&&+\sum_{\mathbf{k}} \sum_{\alpha \neq 0, \beta \neq 0}\left( B_{\mathbf{k}\alpha \beta }b_{%
\mathbf{k}\alpha }^{\dagger }b_{\mathbf{-k}\beta }^{\dagger }+B_{\mathbf{k}%
\alpha \beta }b_{\mathbf{-k}\beta }b_{\mathbf{k}\alpha }\right)\notag
.
\end{eqnarray}
The specific form of $H^{(0)}$ and the matrices $A_{\mathbf{k}\alpha \beta}$ and $B_{\mathbf{k}%
\alpha \beta }$ can be found in the SM.

The quadratic mean-field Hamiltonian $\left( \ref{MF}\right) $ can be
diagonalized by means of a Bogoliubov transformation,  $c_{\mathbf{k}\alpha }^{\dagger }=\sum_{\beta }X_{\mathbf{k}\beta \alpha }b_{%
\mathbf{k}\beta }^{\dagger }-\sum_{\beta}Y_{\mathbf{k}\beta
\alpha }b_{\mathbf{-k}\beta}$, leading to the Bogoliubov eigensystem equation \cite{Ripka}
\begin{equation}
\left(
\begin{array}{cc}
A_{\mathbf{k}} & 2B_{\mathbf{k}} \\
-2B_{\mathbf{k}}^{\ast } & -A_{\mathbf{k}}^{\ast }%
\end{array}%
\right) \left(
\begin{array}{c}
X_{\mathbf{k}} \\
Y_{\mathbf{k}}%
\end{array}%
\right) =\omega _{\mathbf{k}}\left(
\begin{array}{c}
X_{\mathbf{k}} \\
Y_{\mathbf{k}}%
\end{array}%
\right) ,
\label{RPA}
\end{equation}%
where the positive eigenvalues $\omega _{\mathbf{k}}$
determine the excitation spectrum.
The Bogoliubov equation depends on the
generic transformation $U$ previously defined.
Upon minimization of the free energy with respect
to the condensed CB structure $U_{\mathbf{m}}^{0}$
we derive a Hartree like equation for this
transformation. The resulting equation can be cast in matrix form
\begin{equation}
\sum_{\mathbf{n}}h_{\mathbf{m,n}}U_{\mathbf{n}}^{0}=\lambda U_{\mathbf{m}%
}^{0}.  \label{h}
\end{equation}%

The derivation of the matrix elements of $h$ for $L \times L$ clusters, given in the SM, is straightforward though lengthy.
The Hartree Hamiltonian $h$ depends on the unitary
transformation $U$, on the Bogoliubov amplitudes
$X_{\mathbf{k}},~~Y_{\mathbf{k}}$, and on the fraction of the
condensate $\sigma^2$. Strictly speaking, the self-consistent
Hartree diagonalization provides a single eigenvector defining the
structure of the condensed CB and the
corresponding lowest eigenvalue, which is the Lagrange multiplier
$\lambda $. However, after attaining self-consistency the matrix
diagonalization procedure supplies a complete set of eigenstates
that are orthogonal to the condensed CB. It is in this basis
orthogonal to the condensate where the mean-field Hamiltonian
(\ref{MF}) is expressed.
 We seek a
self-consistent solution of the coupled set of equations given by the
Hartree eigensystem (\ref{h}) which fixes the unitary
transformation $U$ and the Langrange multiplier $\lambda$, the Bogoliubov equations (\ref{RPA}) that provide the Bogoliubov amplitudes
$X_{\mathbf{k}}$ and $Y_{\mathbf{k}}$, together with the expectation value of the physical constraint ($\ref{Constraint}$)
 that determines the CB condensed fraction $\sigma ^{2}$.

{\it 2D Bose-Hubbard Model Results.}-- We start with benchmark calculations based on $2\times 2$ clusters describing the first Mott lobe
characterized by a fixed density per site $\rho=1$.
Within this phase, the structure of the unitary transformation $U$ and the CB condensed fraction $
\sigma^{2}$ are $\mu$-independent. The structure of the condensed
CB, dictated by $U^0$, is a linear combination of cluster states with $\vert\mathbf{n}\vert=4$. The relevant CB fluctuations are pairs of \textit{particle} $\left(\vert\mathbf{n}\vert=5\right)$ and
\textit{hole} $\left(\vert\mathbf{n}\vert=3\right)$ cluster states.
In addition, particle- and hole-like excitation eigenvalues have a linear
dependence on the chemical potential for fixed $t$. Both excitations cross each other at the \textit{%
p-h symmetry line}, where the gap is doubly degenerate. The edges of the first Mott lobe are determined by the vanishing of the gap, indicating the appearance of a Goldstone
mode at $\mathbf{k=0}$ related to the $U(1)$ symmetry
breaking in the superfluid.
\begin{figure}[t]
\includegraphics[width=9.5cm]{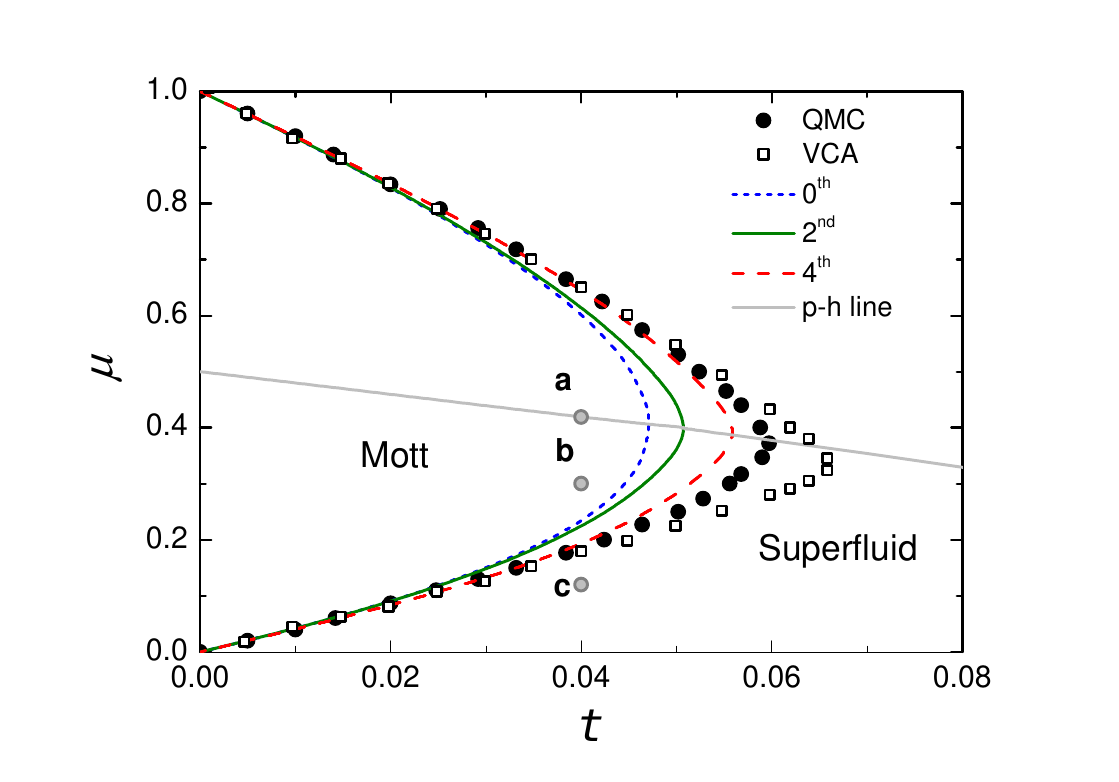}
\caption{(color online) Phase diagram of the first lobe of the Mott-insulator to
superfluid transition. The dotted, solid, and dashed curves show the
$2\times2$ CBMFT results in increasing order of approximation.
The p-h symmetry line traverses the Mott lobe and extends into the superfluid region at a constant density $\rho=1$.
Black circles are QMC results from Ref. \cite{QMC}. Squares display VCA results from Ref. \cite{KAL1}. Grey circles display three points in
parameter space for which the dispersions are analyzed in Fig. 3.}
\label{Diag}
\end{figure}

Fig. 1 shows the phase diagram of the Bose-Hubbard model in three different CB mean-field approximations. The $0^{th}$ order approximation neglects fluctuations and solves the Hartree equations exclusively. The edges of the Mott lobe are determined in this case by a deviation from the density $\rho=1$. This $0^{th}$ order approximation is equivalent to the cluster mean-field calculations of Refs. \cite{Pek,Multisite}  producing the same phase diagram (dotted line in Fig. 1).  The $2^{nd}$ and $4^{th}$ order approximations go beyond previous cluster mean-field approximations incorporating fluctuations by means of a self-consistent solution of the Bogoliubov (\ref{RPA}) plus Hartree (\ref{h}) equations linked by the physical condition (\ref{Constraint}). The $2^{nd}$ order approximation neglects two-body interactions among fluctuating bosons, while the $4^{th}$ order solves the three coupled equations in full. As the approximation order increases, CBMFT shows clear convergence towards QMC. VCA results, which were related in Ref. \cite{KAL1} to a linear approximate CB mapping, extend well beyond the QMC Mott lobe. Also shown in Fig. 1 is the extension of the p-h line into the superfluid phase characterized by density $\rho=1$.

The full self-consistent $4^{th}$ order approximation does not describe the gapless feature of the
superfluid phase correctly. Although ways to correct this deficiency have been suggested \cite{Yuka}, in the
rest of this paper we will focus on the $2^{nd}$ order approximation that strictly preserves the gapless spectrum when $U(1)$ symmetry is broken.

\begin{figure}[t]

\includegraphics[width=9.5cm]{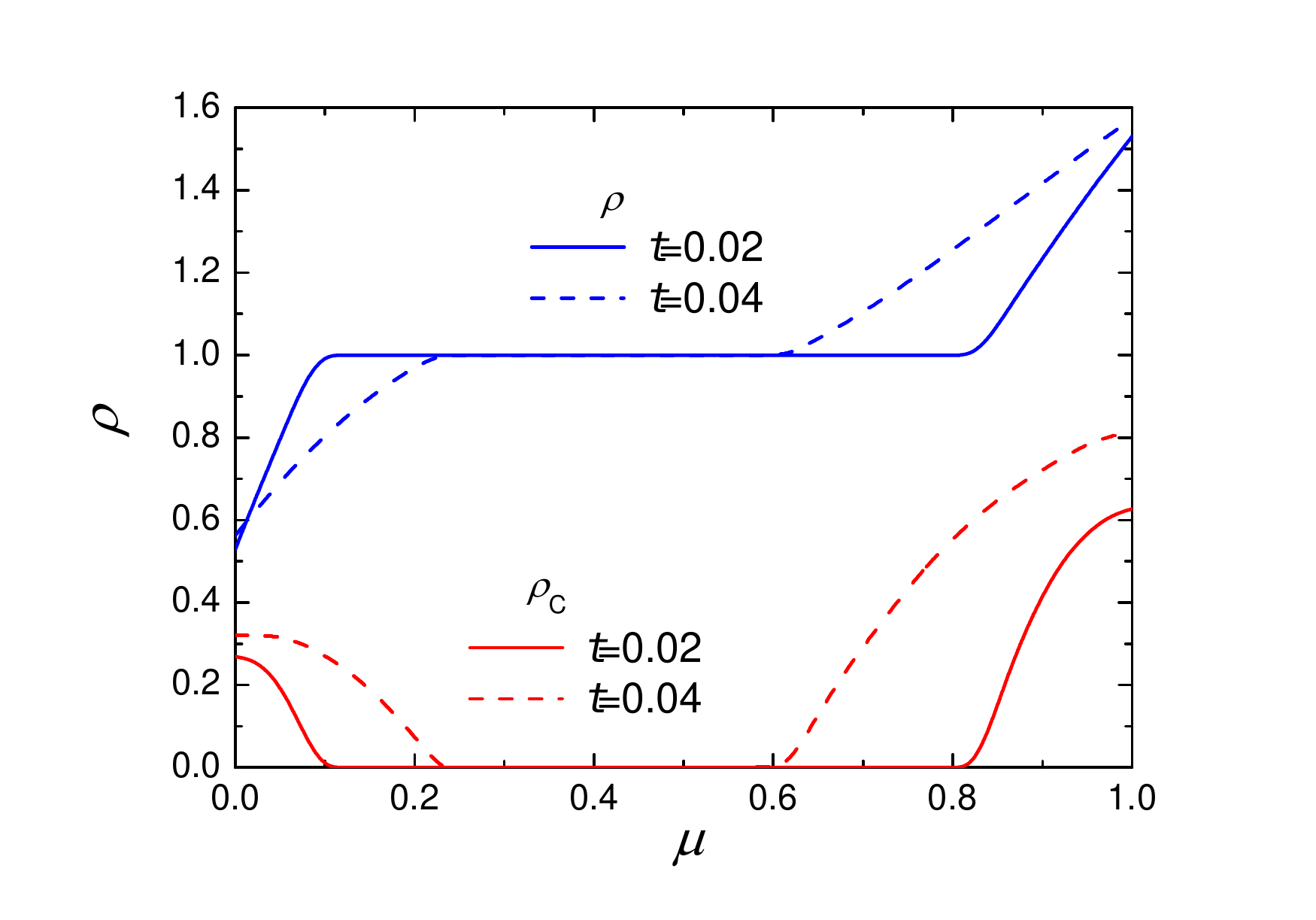}
\caption{(color online). Total density ($\protect\rho$) and condensate density ($\rho
_{c}$) for $t=0.02$ (solid line) and $t=0.04$ (dashed line) within the
second order approximation.  }
\label{Dens}
\end{figure}
Fig. 2 shows the total density $\rho=\langle \phi\vert a_{j}^{\dagger}a_{j}\vert \phi\rangle$ and the condensate density  $\rho_{c}=\vert
\langle \phi\vert a_{j}^{\dagger} \vert \phi\rangle \vert ^{2}$ for
 hopping values of  $t=0.02$ and $0.04$.  The plateau characterizing the Mott phase is reduced
 for larger $t$. Outside this region, the superfluid has non-commensurate
density.
The condensate density of physical bosons, representing the
coherence of the superfluid phase, vanishes in the Mott phase. VCA results for $t=0.02$ \cite{KAL1,KAL2} compare well with
 our results.
\begin{figure}[t]
\includegraphics[width=9.5cm]{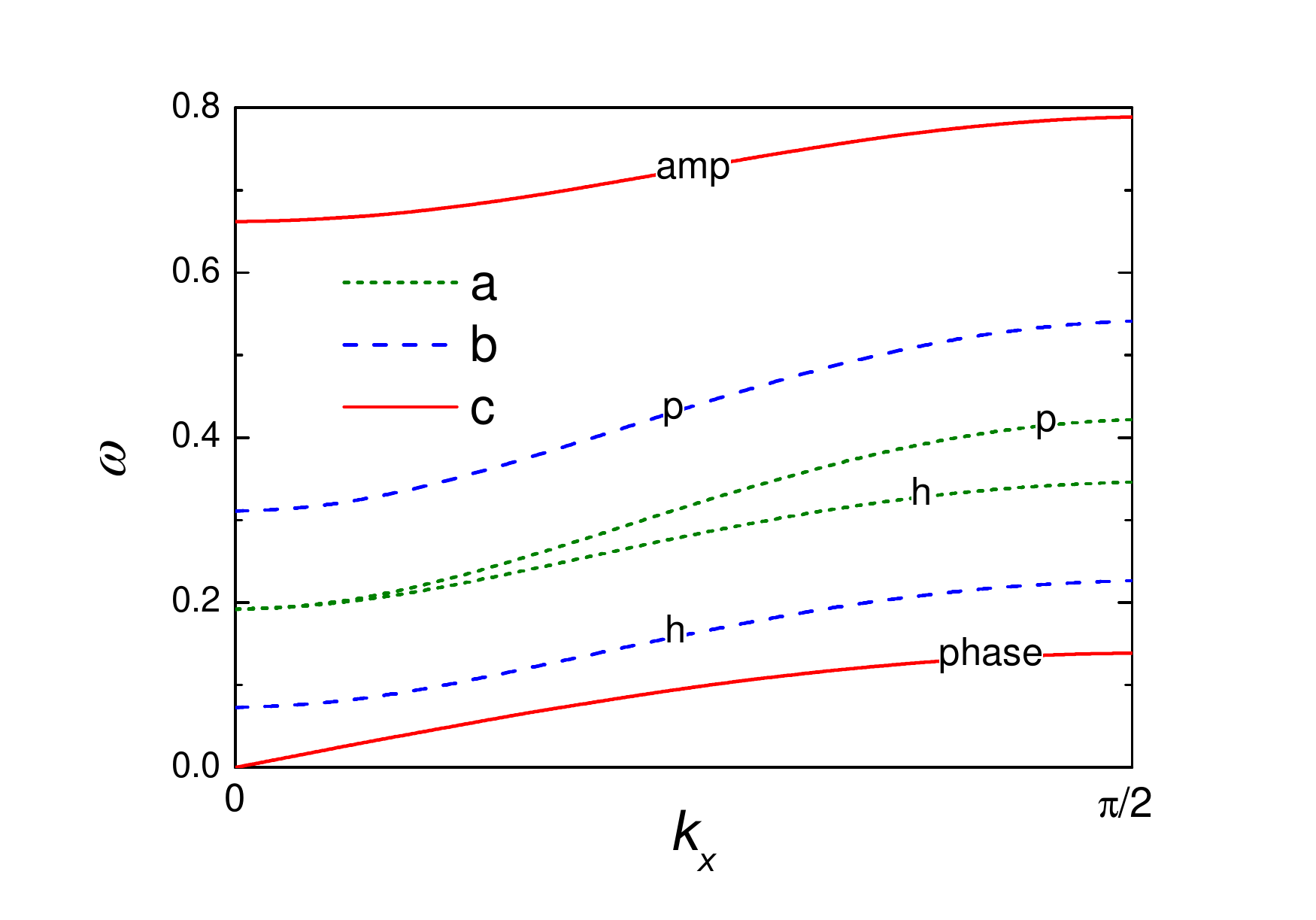}
\caption{(color online). Dispersion modes for $t=0.04$:  particle- and hole-like excitation modes  at ({\bf a}) the p-h symmetry line($\protect\mu=0.419$), and ({\bf b}) inside the
Mott insulator ($\protect\mu=0.30$), and amplitude- and phase-like modes at ({\bf c}) in the superfluid ($\protect\mu=0.12$).}
\label{spectra}
\end{figure}
\begin{figure}[t]
\includegraphics[width=9.5cm]{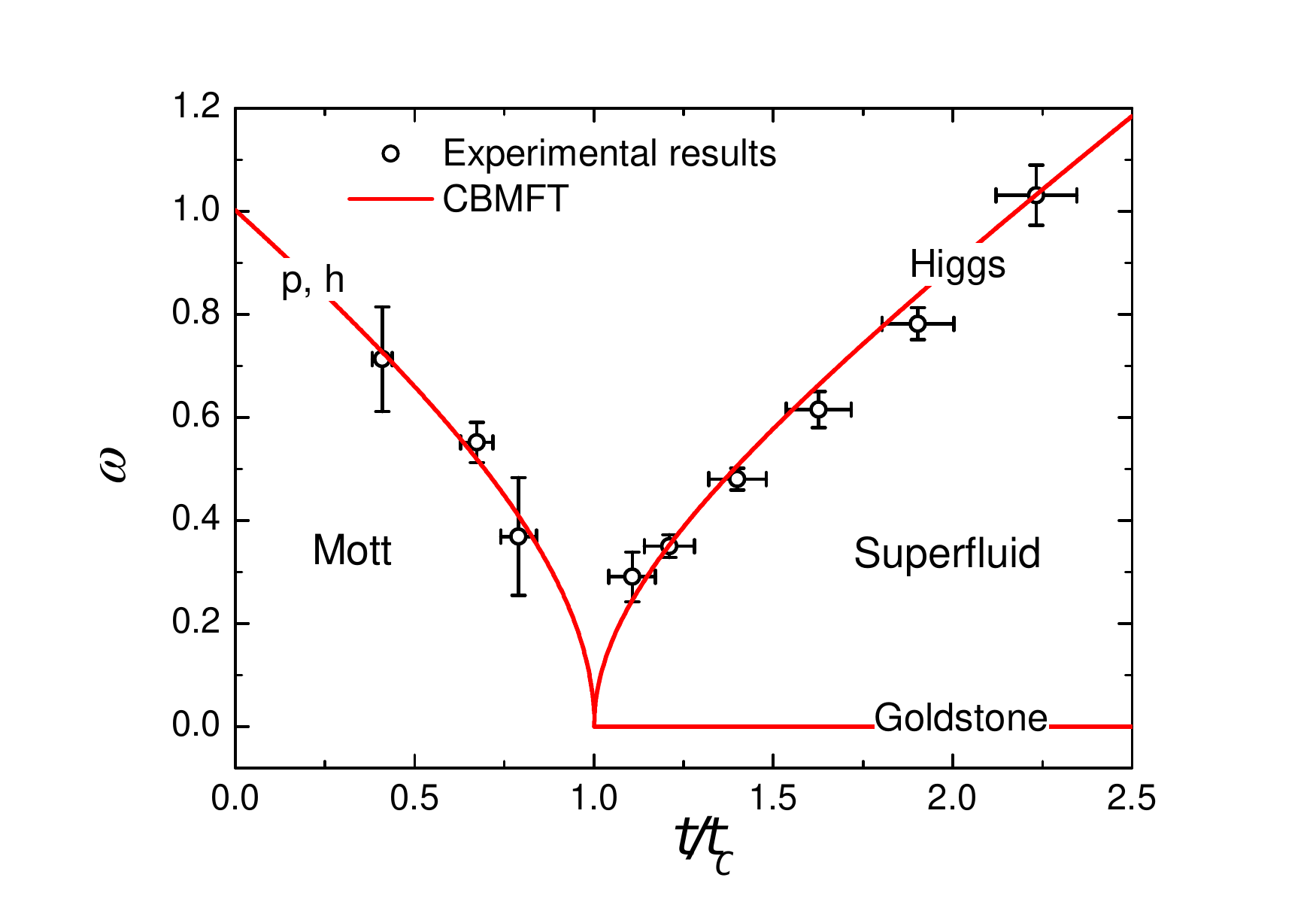}
\caption{(color online). Higgs, Goldstone, particle, and hole modes along the p-h symmetry line computed within
$2^{nd}$ order CBMFT (solid line). Experimental data points from \protect\cite{Endres}.}
\label{higgs}
\end{figure}

In Fig. 1, we have depicted three characteristic points at $t=0.04$; namely, ${\bf a}$ is at the p-h line in the Mott phase, ${\bf b}$ is still in the Mott phase but away from the p-h line, and ${\bf c}$ is in the superfluid phase.
Fig. 3 shows particle- and hole-like excitations for $k_y=0$ as a function of $k_x$ for ${\bf a}$ and ${\bf b}$ inside the Mott phase.
The degeneracy of the particle and hole modes for point ${\bf a}$ approaching $\mathbf{k=0}$ is clearly seen in this figure. Away from the p-h line and still in the Mott phase (point ${\bf b}$), this degeneracy is broken and the hole is favored against the particle mode. Well inside the superfluid phase (point ${\bf c}$), we recognize a gapless mode (Goldstone) with the characteristic linear dispersion at low momentum, as well as a gapped mode. An analysis of the CB structure $U^{\alpha}_{\mathbf{m}}$ of each mode, similar to the one performed in Ref. \cite{Pek} shows that the gapless mode is a phase-like mode, while the gapped mode is an amplitude-like mode.

The phase transition taking place at the lobe tip along a constant density line $\left(\rho=1\right)$ can be understood in terms of an $O(2)$ relativistic field theory, as has been recently discussed in Refs. \cite{Pollet, Endres}.
Fig. \ref{higgs} displays how doubly degenerate excitations along the p-h line inside the Mott insulator
vanish at the critical point. In the superfluid region, one of them remains at zero excitation energy (Goldstone)
while the other one grows for increasing hopping (Higgs). In both cases, their structure mixes particle- and hole-like states of the cluster.  The CBMFT results not only match the experimental data \cite{Endres} remarkably well but also gives an excellent description of the critical point.

{\it Conclusions.}-- We have introduced a cluster composite boson mapping which separates intra-
and inter-cluster degrees of freedom. The former are treated exactly while
the latter can be approximated using standard many-body methods applied
to the resulting CB Hamiltonian. We have here shown
that a mean-field approximation to the CB interaction for the Bose-Hubbard model
produces an accurate description of the Mott-superfluid phase diagram compared
to QMC results. Densities and dispersions are found in
quantitative agreement with more sophisticated techniques like VCA.
The recently measured Higgs mode is also computed and found to be
in remarkable agreement with experiment. Further improvement of the theory
beyond the mean-field $2^{nd}$ order approximation employed in this work
is feasible. Most importantly, CBMFT is readily applicable to
other many-body problems where frustration, synthetic gauge fields or long range interactions
pose significant hurdles to existing state-of-the-art methodologies.


We acknowledge useful discussions with C A. Jimenez-Hoyos, L. Isaev, and G.
Ortiz. This work was supported by grants
FIS2009-07277, FIS2012-34479 and BES-2010-031607 of the Spanish Ministry of Economy and Competitiveness. GES is supported by DOE DE-FG02-09ER16053 and The Welch Foundation (C-0036).

\end{document}


\onecolumngrid
\title{Supplemental Material}
\maketitle
\section{The Composite Boson Mapping}

In this section we demonstrate that the CB mapping satisfies the commutation relations of the original boson operators $a_i, a^{\dag}_j$. We start with the commutator $\left[a_{i},a_{j}\right]$ when $i,j\in R$. Making use of the quadratic mapping in Eq. (1), the commutator maps to

\begin{eqnarray}
\left[a_i,a_j\right]&=& \sum_{\alpha,\beta}\sum_{\alpha',\beta'}
\langle R\alpha\vert a_i\vert R\beta\rangle\langle R\alpha'\vert a_j\vert R\beta'\rangle
\left(b_{R\alpha}^{\dagger}b_{R\beta}b^{\dagger}_{R\alpha'}b_{R\beta'}-b_{R\alpha'}^{\dagger}b_{R\beta'}b^{\dagger}_{R\alpha}b_{R\beta}\right).
\end{eqnarray}

Applying the canonical commutation relations of the CB bosons,
$\left[b_{R\alpha},b^{\dag}_{R' \beta}\right]=\delta_{\alpha\beta}\delta_{RR'}$, we arrive to
\begin{eqnarray}
\left[a_i,a_j\right]&=&\sum_{\alpha,\beta}\sum_{\alpha',\beta'}
\langle R\alpha\vert a_i\vert R\beta\rangle\langle R\alpha'\vert a_j\vert R\beta'\rangle
\left(
b_{R\alpha}^{\dagger}\left[ \delta_{\alpha',\beta}+b^{\dagger}_{R\alpha'}b_{R\beta} \right]b_{R\beta'}
-b_{R\alpha'}^{\dagger}\left[ \delta_{\alpha,\beta'}+b^{\dagger}_{R\alpha}b_{R\beta'} \right]b_{R\beta}
\right),
\end{eqnarray}

and using  the resolution of the identity, $\sum_{\alpha}\vert R\alpha\rangle\langle R\alpha\vert=1$,
\begin{eqnarray}
\left[a_i,a_j\right]=\sum_{\alpha,\beta}
\langle\alpha\vert a_{i}a_{j}\vert\beta\rangle b^{\dagger}_{R\alpha}b_{R\beta}
-\sum_{\alpha\beta}\langle R\alpha\vert a_{j}a_{i}\vert R\beta\rangle b^{\dagger}_{R\alpha}b_{R\beta}
=\sum_{\alpha\beta}\langle\alpha\vert \left[a_{i},a_{j}\right]\vert \beta\rangle b_{R\alpha}^{\dagger}b_{R\beta}=0.
\end{eqnarray}

When $i\in R$ and $j\in R'$ the commutator trivially vanishes, as the CB operators commute for different clusters.

Equivalently, we show that the CB images of the elementary boson operators preserve the boson commutation relation,
$\left[a_i,a^{\dag}_j\right]=\delta_{ij}$. Making use of Eq. (1)

\begin{eqnarray}
\left[a_i,a^{\dag}_j\right]&=& \sum_{\alpha,\beta}\sum_{\alpha',\beta'}
\langle R\alpha\vert a_i\vert  R\beta\rangle\langle R'\alpha'\vert a^{\dag}_j\vert R'\beta'\rangle
\left(b_{R\alpha}^{\dagger}b_{R\beta}b^{\dagger}_{R'\alpha'}b_{R'\beta'}-b_{R'\alpha'}^{\dagger}b_{R'\beta'}b^{\dagger}_{R\alpha}b_{R\beta}\right).
\end{eqnarray}

Using the commutation relations of the CB bosons
\begin{eqnarray}
\left[a_i,a^{\dag}_j\right]&=&\sum_{\alpha,\beta}\sum_{\alpha',\beta'}
\langle R\alpha\vert a_i\vert R\beta\rangle\langle R'\alpha'\vert a^{\dag}_j\vert R'\beta'\rangle\notag\\
&&\times
\left(
b_{R\alpha}^{\dagger}\left[ \delta_{\alpha',\beta}\delta_{RR'}+b^{\dagger}_{R'\alpha'}b_{R\beta} \right]b_{R'\beta'}
-b_{R'\alpha'}^{\dagger}\left[ \delta_{\alpha,\beta'}\delta_{RR'}+b^{\dagger}_{R\alpha}b_{R'\beta'} \right]b_{R\beta}
\right),
\end{eqnarray}

using  the resolution of the identity, $\sum_{\alpha}\vert R\alpha\rangle\langle R\alpha\vert=1$,
\begin{eqnarray}
\left[a_i,a^{\dag}_j\right]=\sum_{\alpha,\beta}
\langle R\alpha\vert a_{i}a^{\dag}_{j}\vert R\beta\rangle b^{\dagger}_{R\alpha}b_{R\beta}
-\sum_{\alpha\beta}\langle R\alpha\vert a^{\dag}_{j}a_{i}\vert R\beta\rangle b^{\dagger}_{R\alpha}b_{R\beta}
=\sum_{\alpha\beta}\langle R\alpha\vert \left[a_{i},a^{\dag}_{j}\right]\vert R\beta\rangle b_{R\alpha}^{\dagger}b_{R\beta},
\end{eqnarray}

and using orthonormality of the CB basis and the physical constraint $\sum_{\alpha}b_{R\alpha}^{\dagger}b_{R\alpha}=1$,
\begin{eqnarray}
\left[a_i,a^{\dag}_j\right]=\delta_{ij} \sum_{\alpha}b_{R\alpha}^{\dagger}b_{R\alpha}=\delta_{ij}.
\end{eqnarray}

We have demonstrated that the mapping is canonical if the resolution of the identity and the physical constraint are satisfied.

\bigskip
Let us now formally show that the mapping of the kinetic and interaction terms of the Bose-Hubbard Hamiltonian inside the cluster lead to a one-body CB term. Following the previous procedure, we map the hopping operator $a_{i}^{\dagger}a_{j}$ acting inside a cluster, $i,j\in R$. Using the mapping of Eq. (1) we obtain a two-body CB operator
\begin{eqnarray}
a_{i}^{\dagger}a_{j}&=&\sum_{\alpha\beta}\sum_{\alpha'\beta'}
\langle R\alpha\vert a_{i}^{\dagger}\vert R\beta\rangle\langle R\alpha'\vert a_{j}\vert R\beta'\rangle
b_{R\alpha}^{\dagger}b_{R\beta}b_{R\alpha'}^{\dagger}b_{R\beta'},
\end{eqnarray}

using the commutation relations of CB bosons
\begin{eqnarray}
a_{i}^{\dagger}a_{j}&=&
\sum_{\alpha\beta}\sum_{\alpha'\beta'}
\langle R\alpha\vert a_{i}^{\dagger}\vert R\beta\rangle\langle R\alpha'\vert a_{j}\vert R\beta'\rangle
b_{R\alpha}^{\dagger}\left[\delta_{\alpha'\beta}+b_{R\alpha'}^{\dagger}b_{R\beta}\right]b_{R\beta'},
\end{eqnarray}

and applying  the resolution of the identity and the physical constraint we finally arrive to

\begin{eqnarray}
a_{i}^{\dagger}a_{j}&=&\sum_{\alpha\beta}\langle R\alpha\vert a_{i}^{\dagger}a_{j}\vert R\beta\rangle b_{R\alpha}^{\dagger}b_{R\beta}
\end{eqnarray}

Next, we can map the interaction term within the cluster, $n_i n_j=a_{i}^{\dag}a_{i}a_{j}^{\dag}a_{j}$, $i,j\in R$. First, we apply the mapping of Eq. (1) arriving to a four-body operator in the CB space
\begin{eqnarray}
n_{i}n_{j}
&=&\sum_{\alpha_{1}\beta_{1}}\sum_{\alpha_{2}\beta_{2}}\sum_{\alpha_{3}\beta_{3}}\sum_{\alpha_{4}\beta_{4}}
\langle R\alpha_{1}\vert a_{i}^{\dagger}\vert R\beta_{1}\rangle
\langle R\alpha_{2}\vert a_{i}\vert R\beta_{2}\rangle
\langle R\alpha_{3}\vert a_{j}^{\dagger}\vert R\beta_{3}\rangle
\langle R\alpha_{4}\vert a_{j}\vert R\beta_{4}\rangle\notag\\
&&~~~~~~~~~~~~~~~~~~~~\times
b^{\dagger}_{R\alpha_{1}}b_{R\beta_{1}}
b^{\dagger}_{R\alpha_{2}}b_{R\beta_{2}}
b^{\dagger}_{R\alpha_{3}}b_{R\beta_{3}}
b^{\dagger}_{R\alpha_{4}}b_{R\beta_{4}}.
\end{eqnarray}

Then, we commute two CB operators

\begin{eqnarray}
n_{i}n_{j}&=&
\sum_{\alpha_{1}\beta_{1}}\sum_{\alpha_{2}\beta_{2}}\sum_{\alpha_{3}\beta_{3}}\sum_{\alpha_{4}\beta_{4}}
\langle R\alpha_{1}\vert a_{i}^{\dagger}\vert R\beta_{1}\rangle
\langle R\alpha_{2}\vert a_{i}\vert R\beta_{2}\rangle
\langle R\alpha_{3}\vert a_{j}^{\dagger}\vert R\beta_{3}\rangle
\langle R\alpha_{4}\vert a_{j}\vert R\beta_{4}\rangle\notag\\
&&~~~~~~~~~~~~~~~~~~~~\times
b^{\dagger}_{R\alpha_{1}}b_{R\beta_{1}}
b^{\dagger}_{R\alpha_{2}}
\left[\delta_{\alpha_{3}\beta_{2}}+b^{\dagger}_{R\alpha_{3}}b_{R\beta_{2}}\right]
b_{R\beta_{3}}
b^{\dagger}_{R\alpha_{4}}b_{R\beta_{4}},
\end{eqnarray}

and we apply resolution of the identity and the physical constraint
\begin{eqnarray}
n_{i}n_{j}&=&
\sum_{\alpha_{1}\beta_{1}}\sum_{\alpha_{2}\beta_{3}}\sum_{\alpha_{4}\beta_{4}}
\langle R\alpha_{1}\vert a_{i}^{\dagger}\vert R\beta_{1}\rangle
\langle R\alpha_{2}\vert
a_{i}
a_{j}^{\dagger}\vert R\beta_{3}\rangle
\langle R\alpha_{4}\vert a_{j}\vert R\beta_{4}\rangle
b^{\dagger}_{R\alpha_{1}}b_{R\beta_{1}}
b^{\dagger}_{R\alpha_{2}}
b_{R\beta_{3}}
b^{\dagger}_{R\alpha_{4}}b_{R\beta_{4}}.
\end{eqnarray}

Following the previous procedure repeatedly (use of commutation relations of the CB operators
and application of the resolution of identity and the physical constraint) we arrive to the final result

\begin{equation}
n_{i}n_{j}=
\sum_{\alpha\beta}
\langle R\alpha\vert a_{i}^{\dagger}
a_{i}
a_{j}^{\dagger}
a_{j}\vert R\beta\rangle
b^{\dagger}_{R\alpha}
b_{R\beta}.
\end{equation}

The hopping operator term $a_{i}^{\dagger}a_{j}$ transferring particles from one cluster to a neighbor cluster, $i\in R$ and $j\in R'\neq R$, maps to a two-body CB term. Applying the mapping and normal ordering

\begin{eqnarray}
a_{i}^{\dagger}a_{j}
&=&\sum_{\alpha\beta}\sum_{\alpha'\beta'}
\langle R\alpha\vert\langle R'\alpha'\vert a_{i}^{\dagger}a_{j}\vert R\beta\rangle\vert R'\beta'\rangle
~b_{R\alpha}^{\dagger}b^{\dagger}_{R'\alpha'}b_{R\beta}b_{R'\beta'}.
\end{eqnarray}

Last, we map the interaction term $n_{i}n_{j}$ between two cluster
$i\in R$ and $j\in R'\neq R$,  leading to a four-body CB term

\begin{eqnarray}
n_{i}n_{j}
&=&\sum_{\alpha_{1}\beta_{1}}\sum_{\alpha_{2}\beta_{2}}\sum_{\alpha_{3}\beta_{3}}\sum_{\alpha_{4}\beta_{4}}
\langle R\alpha_{1}\vert a_{i}^{\dagger}\vert R\beta_{1}\rangle
\langle R\alpha_{2}\vert a_{i}\vert R\beta_{2}\rangle
\langle R'\alpha_{3}\vert a_{j}^{\dagger}\vert R'\beta_{3}\rangle
\langle R'\alpha_{4}\vert a_{j}\vert R'\beta_{4}\rangle\notag\\
&&~~~~~~~~~~~~~~~~~~~~\times b^{\dagger}_{R\alpha_{1}}b_{R\beta_{1}}
b^{\dagger}_{R\alpha_{2}}b_{R\beta_{2}}
b^{\dagger}_{R'\alpha_{3}}b_{R'\beta_{3}}
b^{\dagger}_{R'\alpha_{4}}b_{R'\beta_{4}}.
\end{eqnarray}

Following a similar procedure of commutation, resolution of the identity and application of the physical constraint, we finally arrive to a two-body CB interaction

\begin{eqnarray}
n_{i}n_{j}
&=&\sum_{\alpha\beta}\sum_{\alpha'\beta'}
\langle R\alpha\vert \langle R'\alpha'\vert
a_{i}^{\dagger}
 a_{i}
 a_{j}^{\dagger}
 a_{j}\vert R\beta\rangle\vert R'\beta'\rangle
b^{\dagger}_{R\alpha}
b^{\dagger}_{R'\alpha'}
b_{R\beta}
b_{R'\beta'}.
\end{eqnarray}

We have demonstrated that both the hopping operator and the density-density interaction within a cluster map to a one-body CB operator. Furthermore,  we showed that when these operators act between neighbor clusters they are mapped into two-body CB operators.

\bigskip

\section{Matrix Elements of the CB Hamiltonian}

The interaction between two clusters appearing in Eq. (5) depends on the particular layout of the two clusters $R$ and $R'$. For the Bose-Hubbard model, it is exclusively related to the hopping term that destroys (creates) a boson at the edge of one cluster and creates (destroys) a boson at the edge of a neighbor cluster. The specific form of the interaction for $ L\times L$ clusters is

\begin{eqnarray}
W^{\alpha\alpha'}_{\beta\beta'}(R,R')&=&\sum_{\langle i,j\rangle}\sum_{\mathbf{n,n'}}
\left(\sqrt{n_i+1}\sqrt{n_j^{\prime}}~
U^{\alpha}_{R\lbrace n_{i}+1\rbrace}
U^{\alpha'}_{R'\lbrace n_{j}'-1\rbrace}
U^{\beta}_{R\mathbf{n}}
U^{\beta'}_{R'\mathbf{n'}}
+\sqrt{n_i}\sqrt{n_j^{\prime}+1}~
U^{\alpha}_{R\lbrace n_{i}-1\rbrace}
U^{\alpha'}_{R'\lbrace n_{j}'+1\rbrace}
U^{\beta}_{R\mathbf{n}}
U^{\beta'}_{R'\mathbf{n'}}
\right)\label{interaction},\notag\\
\end{eqnarray}
where $\sum_{\langle ij\rangle}$ runs over the $L$ links between the edges of both clusters with $i\in R$ and $j\in R'$.
Using compact notation, $\lbrace n_{i}\pm 1\rbrace$ corresponds to a configuration $\mathbf{n}$ with one more or one less boson at site $i$, that is, $\lbrace n_{i}\pm 1\rbrace\equiv\left(n_1,\ldots,n_{i}\pm1,\ldots,n_{L^2} \right)$.

The intra-cluster one-body matrix elements has both contributions, the hopping term and the Hubbard interaction,

\begin{eqnarray}
T^{\alpha}_{\beta}(R)&=&\sum_{\mathbf{n}} \sum_{j\in R} \left( \frac{1}{2}n_{j}\left(n_{j}-1\right)-\mu n_{j}\right)
U^{\alpha}_{R\mathbf{n}}U^{\beta}_{R\mathbf{n}}\notag\\
&&-t \sum_{\mathbf{n}} \sum_{\langle ij\rangle\in R} \left(
\sqrt{n_{i}+1}\sqrt{n_{j}}~U^{\alpha}_{R\lbrace n_{i}+1,n_{j}-1\rbrace}U^{\beta}_{R\mathbf{n}}
+\sqrt{n_{i}}\sqrt{n_{j}+1}~U^{\alpha}_{R\lbrace n_{i}-1,n_{j}+1\rbrace}U^{\beta}_{R\mathbf{n}}
\right).\label{T}
\end{eqnarray}

For a uniform and translational invariant system in the cluster superlattice, the unitary transformation $U$ does not depend the on the cluster position $R$.
Therefore, we drop the position indices in the matrix elements $T$ and $W$, as it is done in Eq. (6).

\bigskip
\section{Bogoliubov matrix}

After shifting the $b^{(\dag)}_{\alpha=0,\mathbf{k=0}}$ CB,
we can classify the resulting terms of the CB Hamiltonian by the number of fluctuating CBs
$(\alpha \neq 0)$,

\begin{eqnarray}
H^{(0)}&=&2M W_{00}^{00}\sigma ^{4}+ M\left( T_{0}^{0}-\lambda \right) \sigma ^{2}\notag\\
H^{(2)}&=&
\sum_{\mathbf{k}}\sum_{\alpha\beta\neq 0}\left( T_{\beta }^{\alpha}-\lambda \delta _{\alpha \beta}\right)
b^{\dag}_{\mathbf{k}\alpha}b_{\mathbf{k}\beta}
+2\sigma ^{2}\sum_{\mathbf{k}}\sum_{\alpha\beta\neq 0}
\left( W_{0\beta}^{\alpha 0}\gamma_{\mathbf{k}} + 2W_{\beta 0}^{\alpha 0} \right)b^{\dag}_{\mathbf{k}\alpha}b_{\mathbf{k}\beta}\notag\\
H^{(4)}&=&\frac{1}{M}\sum_{\substack{ \alpha \alpha ^{\prime }\beta \beta ^{\prime }\neq 0
}}W_{\beta \beta ^{\prime }}^{\alpha \alpha ^{\prime }}\sum_{\substack{
\mathbf{k}_{1}\mathbf{k}_{2}\mathbf{q}}}\gamma _{\mathbf{q}}~b_{\mathbf{k_{1}%
}\alpha }^{\dagger }b_{\mathbf{k_{2}+q}\alpha ^{\prime }}^{\dagger }b_{%
\mathbf{k_{1}+q}\beta }b_{\mathbf{k_{2}}\beta ^{\prime }}  \notag
\end{eqnarray}

Notice that $H^{(0)}$ accounts for the free-term of the CB condensate and its self-interaction, $H^{(2)}$ accounts for the interaction of the
fluctuations with the condensate, and $H^{(4)}$ accounts for the interaction among fluctuations. From the three orders of approximation discussed and used in the letter, the $0th$ order corresponds to neglect $H^{(2)}$ and $H^{(4)}$, the $2nd$ order corresponds to neglect $H^{(4)}$, and the $4th$ order takes all terms into account. 
The $H^{(4)}$ term is mean-field decoupled using Wick's theorem in order to obtain the quadratic mean-field Hamiltonian of Eq. (8), where the matrices $A_{\mathbf{k}}$ and $B_{\mathbf{k}}$ are given by
\begin{eqnarray}
A_{\mathbf{k}\alpha\beta}&=&
\left( T_{\beta }^{\alpha}-\lambda \delta _{\alpha \beta}\right)
+2\sigma ^{2}
\left( W_{0\beta}^{\alpha 0}\gamma_{\mathbf{k}} + 2W_{\beta 0}^{\alpha 0} \right)
+\frac{2}{M}\sum_{\mathbf{q}}\sum_{\alpha'\beta'}
\left( W^{\alpha \alpha' }_{\beta'\beta}
\Gamma_{\mathbf{kq}}
+ 2W^{\alpha \alpha' }_{\beta \beta'}\right)
P_{\mathbf{q}\alpha'\beta'}
\notag\\
B_{\mathbf{k}\alpha\beta}&=&
\sigma ^{2}W_{00}^{\alpha \beta}\gamma_{\mathbf{k}}
+\frac{1}{M}\sum_{\mathbf{q}}\sum_{\alpha'\beta'}
W^{\alpha \beta }_{\alpha' \beta'}
\Gamma_{\mathbf{kq}}
K_{\mathbf{q\alpha'\beta'}}
\end{eqnarray}

where $\Gamma_{\mathbf{kq}}=\cos(Lk_{x})\cos(Lq_{x})+\cos(Lk_{y})\cos(Lq_{y})$. 
The density matrix $P_{\mathbf{q}\alpha\beta}=\left\langle b_{\mathbf{q}\alpha' }^{\dagger }b_{\mathbf{q}\beta'}\right\rangle=P^{*}_{\mathbf{q}\beta\alpha} $ is hermitian, and the pairing tensor $K_{\mathbf{q}\alpha\beta}=\left\langle b_{\mathbf{q}\alpha' }^{\dagger }b_{-\mathbf{q}\beta'}^{\dagger}\right\rangle
=K_{\mathbf{-q}\beta\alpha}$ is symmetric. For the particular case of the Bose-Hubbard model, we assume that both of them are real. Upon inversion of the
Bogoliubov transformation, it is straightforward to
show that
\begin{equation}
P_{\mathbf{k}\alpha\beta}=\sum_{\alpha'}
Y_{\mathbf{k}\alpha\alpha'}Y_{\mathbf{k}\beta\alpha'},
~~K_{\mathbf{k}\alpha\beta}=\sum_{\alpha'}
Y_{\mathbf{k}\alpha\alpha'}X_{\mathbf{k}\beta\alpha'}.
\end{equation}

The amplitudes of the Bogoliubov transformation $X_{\mathbf{k}}$ and $Y_{\mathbf{k}}$ are the components of the eigenvectors of the Bogoliubov eigensystem (9).
\bigskip
\section{Hartree-Bose matrix}

Minimization of the free energy with respect to the condensed CB structure $U^{(0)}$
leads to the Hartree-Bose (HB) eigensystem (10),

\begin{eqnarray}
\sum_{\mathbf{n}}h_{\mathbf{m,n}}~ U^{0}_{\mathbf{n}}&=&
\sum_{\mathbf{n}}\left( h_{\mathbf{m,n}}^{(0)}+h_{\mathbf{m,n}}^{(2)} \right)U^{0}_{\mathbf{n}}=\lambda U^{0}_{\mathbf{m}},
\end{eqnarray}
where we have made explicit distinction between the the zeroth order
$h^{(0)}$
and
second order $h^{(2)}$ contributions. These come from the derivation of the zeroth,
$\langle H^{(0)}\rangle$, and second, $\langle H^{(2)}\rangle$, order contributions to
the free energy. The zeroth order term
contains the kinetic and interaction terms of the condensate. The second order term contains the interaction of the condensate $\left(\alpha= 0\right)$ with the fluctuations
$\left(\alpha\neq 0\right)$.

\begin{eqnarray}
\langle H^{(0)}\rangle&=& 2M W^{00}_{00}\sigma^{4} +
M \left[ T^{0}_{0} - \lambda\left(\sum_{\mathbf{n}}U^{(0)}_{\mathbf{n}}U^{(0)}_{\mathbf{n}}\right) \right] \sigma^{2},\label{h0}\\
\langle H^{(2)}\rangle&=& -\lambda \sum_{\mathbf{k}}\sum_{\alpha}P_{\mathbf{k}\alpha\alpha}+2\sigma^{2}\sum_{\mathbf{k}}\sum_{\alpha\beta}\left(
 W^{\alpha\beta}_{00} \gamma_{\mathbf{k}} K_{\mathbf{k}\alpha\beta}
+ \left[W^{\alpha 0}_{0\beta} \gamma_{\mathbf{k}}  + 2W^{\alpha 0}_{\beta 0}\right] P_{\mathbf{k}\alpha\beta}
\right).\label{h2}
\end{eqnarray}

Making use of Eqs. (S.19) and (S.20) and assuming $C_{4}$ symmetry of the cluster superlattice,
we write explicitly the tensors entering in Eq. (S.24)

\begin{eqnarray}
W^{00}_{00}&=& \frac{1}{2}\sum_{\langle i,j\rangle}
\left(\sum_{\mathbf{n}}\sqrt{n_{i}+1}~U^{0}_{\lbrace n_{i}+1\rbrace}U^{0}_{\mathbf{n}} \right)
\left(\sum_{\mathbf{n'}}\sqrt{n_{j}'+1}~U^{0}_{\lbrace n_{j}'+1\rbrace}U^{0}_{\mathbf{n'}} \right),\\
T^{0}_{0}&=& \sum_{\mathbf{n}}\sum_{j}\left( \frac{1}{2}n_{j}\left(n_{j}-1\right)-\mu n_{j}\right)
U^{0}_{\mathbf{n}}U^{0}_{\mathbf{n}}\nonumber\\
&& -t\sum_{\mathbf{n}}\sum_{\langle ij\rangle}\left(
\sqrt{n_{i}+1}\sqrt{n_{j}}~ U^{0}_{\lbrace n_{i}+1,n_{j}-1\rbrace}U^{0}_{\mathbf{n}}
+\sqrt{n_{i}}\sqrt{n_{j}+1}~ U^{0}_{\lbrace n_{i}-1,n_{j}+1\rbrace}U^{0}_{\mathbf{n}}
\right).
\end{eqnarray}

The corresponding derivatives with respect to the condensate CB structure $U^{(0)}$ are

\begin{eqnarray}
\frac{\partial}{\partial U^{0}_{\mathbf{m}}}W^{00}_{00}&=&\frac{1}{2}\sum_{\langle i,j\rangle}
\left( \sqrt{m_{i}+1}~U^{0}_{\lbrace m_{i}+1\rbrace} + \sqrt{m_{i}}~U^{0}_{\lbrace m_{i}-1\rbrace} \right)
\left(\sum_{\mathbf{n'}}\sqrt{n_{j}'+1}~U^{0}_{\lbrace n_{j}'+1\rbrace}U^{0}_{\mathbf{n'}} \right)\notag\\
&&+
\frac{1}{2}\sum_{\langle i,j\rangle}
\left(\sum_{\mathbf{n}}\sqrt{n_{i}+1}~U^{0}_{\lbrace n_{i}+1\rbrace}U^{0}_{\mathbf{n}} \right)
\left( \sqrt{m_{j}+1}~U^{0}_{\lbrace m_{j}+1\rbrace} + \sqrt{m_{j}}~U^{0}_{\lbrace m_{j}-1\rbrace} \right)\notag\\
&=&\sum_{\langle i,j\rangle}
\left(\sum_{\mathbf{n}}\sqrt{n_{i}+1}~U^{0}_{\lbrace n_{i}+1\rbrace}U^{0}_{\mathbf{n}} \right)
\left( \sqrt{m_{j}+1}~U^{0}_{\lbrace m_{j}+1\rbrace} + \sqrt{m_{j}}~U^{0}_{\lbrace m_{j}-1\rbrace} \right),
\end{eqnarray}
and
\begin{eqnarray}
\frac{\partial}{\partial U^{0}_{\mathbf{m}}}T^{0}_{0}&=&
2\sum_{j}\left( \frac{1}{2}m_{j}\left(m_{j}-1\right)-\mu m_{j} \right)U^{0}_{\mathbf{n}}\delta_{\mathbf{m,n}}\notag\\
&&- 2t \sum_{\langle ij\rangle} \left(\sqrt{m_{i}} \sqrt{m_{j}+1}~ U^{0}_{\lbrace m_{i}-1,m_{j}+1\rbrace}
+\sqrt{m_{i}+1} \sqrt{m_{j}}~U^{0}_{\lbrace m_{i}+1,m_{j}-1\rbrace}\right).
\end{eqnarray}

Collecting terms, the zeroth order contribution to the HB matrix is

\begin{eqnarray}
h^{(0)}_{\mathbf{m,n}}&=&
\sum_{j}\left( \frac{1}{2}m_{j}\left(m_{j}-1\right)-\mu m_{j} \right)\delta_{\mathbf{m,n}}\notag\\
&&- t \sum_{\langle ij\rangle} \left(\sqrt{m_{i}} \sqrt{m_{j}+1}~ \delta_{\mathbf{n},\lbrace m_{i}-1,m_{j}+1\rbrace}
+\sqrt{m_{i}+1} \sqrt{m_{j}}~\delta_{\mathbf{n},\lbrace m_{i}+1,m_{j}-1\rbrace}\right)\notag\\
&&+ \sigma^{2}\sum_{\langle ij\rangle}
\left[
\left(\sum_{\mathbf{n'}} \sqrt{n_{i}'+1}~U^{0}_{\lbrace n_{i}'+1\rbrace}U^{0}_{\mathbf{n'}}\right)
\left( \sqrt{m_{j}+1}~\delta_{\mathbf{n},\lbrace m_{j}+1\rbrace}+\sqrt{m_{j}}~\delta_{\mathbf{n},\lbrace m_{j}-1\rbrace}
\right)
\right].
\end{eqnarray}

Following a similar procedure, we obtain the second order contributions to the HB matrix (S.23).
First, we write explicitly the $W$ tensors and then their corresponding derivatives with respect to $U^{(0)}$,

\begin{eqnarray}
W^{\alpha\beta}_{00}&=&
\frac{1}{4}\sum_{\langle ij\rangle}\sum_{\mathbf{n,n'}}\left(
\sqrt{n_{i}+1}\sqrt{n_{j}'}~U^{\alpha}_{\lbrace n_{i}+1\rbrace} U^{\beta}_{\lbrace n_{j}'-1\rbrace}U^{0}_{\mathbf{n}}U^{0}_{\mathbf{n'}}
+\sqrt{n_{i}}\sqrt{n_{j}'+1}~U^{\alpha}_{\lbrace n_{i}-1\rbrace} U^{\beta}_{\lbrace n_{j}'+1\rbrace}U^{0}_{\mathbf{n}}U^{0}_{\mathbf{n'}}
\right),\\
\frac{\partial}{\partial U^{0}_{\mathbf{m}}}W^{\alpha\beta}_{00}&=&\frac{1}{4}\sum_{\langle ij\rangle}\sum_{\mathbf{n}}
\left(
\sqrt{n_{i}+1}\sqrt{m_{j}}~U^{\alpha}_{\lbrace n_{i}+1\rbrace}U^{\beta}_{\lbrace m_{j}-1\rbrace}+
\sqrt{n_{i}}\sqrt{m_{j}+1}~U^{\alpha}_{\lbrace n_{i}-1\rbrace}U^{\beta}_{\lbrace m_{j}+1\rbrace}
+\left[n\leftrightarrow m\right]
\right)U^{0}_{\mathbf{n}},
\end{eqnarray}

\begin{eqnarray}
W^{\alpha 0}_{0 \beta}&=&
\frac{1}{4}\sum_{\langle ij\rangle}\sum_{\mathbf{n,n'}}\left(
\sqrt{n_{i}+1}\sqrt{n_{j}'+1}~U^{\alpha}_{\lbrace n_{i}+1\rbrace} U^{0}_{\mathbf{n'}}U^{0}_{\mathbf{n}}U^{\beta}_{\lbrace n_{j}'+1\rbrace}
+\sqrt{n_{i}}\sqrt{n_{j}}~U^{\alpha}_{\lbrace n_{i}-1\rbrace} U^{0}_{\mathbf{n'}}U^{0}_{\mathbf{n}}U^{\beta}_{\lbrace n_{j}'-1\rbrace}
\right),\\
\frac{\partial}{\partial U^{0}_{\mathbf{m}}}W^{\alpha 0}_{0\beta}&=&
\frac{1}{4}\sum_{\langle ij\rangle}\sum_{\mathbf{n}}
\left(
\sqrt{n_{i}+1}\sqrt{m_{j}+1}~U^{\alpha}_{\lbrace n_{i}+1\rbrace}U^{\beta}_{\lbrace m_{j}+1\rbrace}+
\sqrt{n_{i}}\sqrt{m_{j}}~U^{\alpha}_{\lbrace n_{i}-1\rbrace}U^{\beta}_{\lbrace m_{j}-1\rbrace}
+\left[n\leftrightarrow m\right]
\right)U^{0}_{\mathbf{n}},
\end{eqnarray}

\begin{eqnarray}
W^{\alpha 0}_{\beta 0}&=&
\frac{1}{4}\sum_{\langle ij\rangle}
\sum_{\mathbf{n}}\left(\sqrt{n_{i}+1}\left[
U^{\alpha}_{\lbrace n_{i}+1\rbrace}U^{\beta}_{\mathbf{n}}+U^{\alpha}_{\mathbf{n}}U^{\beta}_{\lbrace n_{i}+1\rbrace}
\right]\right)
\left(\sum_{\mathbf{n'}} \sqrt{n_{j}'+1}~U^{0}_{\lbrace n_{j}'+1\rbrace}U^{0}_{\mathbf{n'}}\right),\\
\frac{\partial}{\partial U^{0}_{\mathbf{m}}}W^{\alpha 0}_{\beta 0}&=&\frac{1}{4}\sum_{\langle ij\rangle}
\left(\sum_{\mathbf{n'}}\sqrt{n_{i}'+1}\left[
U^{\alpha}_{\lbrace n_{i}'+1\rbrace}U^{\beta}_{\mathbf{n'}}+U^{\alpha}_{\mathbf{n'}}U^{\beta}_{\lbrace n_{i}'+1\rbrace}
\right]\right)
\left( \sqrt{m_{j}+1}~U^{0}_{\lbrace m_{j}+1\rbrace}+\sqrt{m_{j}}~U^{0}_{\lbrace m_{j}-1\rbrace}
\right).
\end{eqnarray}

Making use of these derivatives, the second order term of the HB matrix is

\begin{eqnarray}
h^{(2)}_{\mathbf{m,n}}&=&
\frac{1}{4 M}\sum_{\langle ij\rangle}\sum_{\alpha\beta}
\left(
\sqrt{n_{i}+1}\sqrt{m_{j}}~U^{\alpha}_{\lbrace n_{i}+1\rbrace}U^{\beta}_{\lbrace m_{j}-1\rbrace}+
\sqrt{n_{i}}\sqrt{m_{j}+1}~U^{\alpha}_{\lbrace n_{i}-1\rbrace}U^{\beta}_{\lbrace m_{j}+1\rbrace}
+\left[n\leftrightarrow m\right]
\right) \left[\sum_{\mathbf{k}}\gamma_{\mathbf{k}}K_{\mathbf{k}\alpha\beta}\right]\notag\\
&+&\frac{1}{4M}\sum_{\langle ij\rangle}\sum_{\alpha\beta}
\left(
\sqrt{n_{i}+1}\sqrt{m_{j}+1}~U^{\alpha}_{\lbrace n_{i}+1\rbrace}U^{\beta}_{\lbrace m_{j}+1\rbrace}+
\sqrt{n_{i}}\sqrt{m_{j}}~U^{\alpha}_{\lbrace n_{i}-1\rbrace}U^{\beta}_{\lbrace m_{j}-1\rbrace}
+\left[n\leftrightarrow m\right]
\right)\left[\sum_{\mathbf{k}}\gamma_{\mathbf{k}}P_{\mathbf{k}\alpha\beta}\right]\\
&+&\frac{1}{2M}\sum_{\langle ij\rangle}\sum_{\alpha\beta}\left[
\left(\sum_{\mathbf{n'}}\sqrt{n_{i}'+1}\left[
U^{\alpha}_{\lbrace n_{i}'+1\rbrace}U^{\beta}_{\mathbf{n'}}+U^{\alpha}_{\mathbf{n'}}U^{\beta}_{\lbrace n_{i}'+1\rbrace}
\right]\right)
\left( \sqrt{m_{j}+1}~\delta_{\mathbf{n},\lbrace m_{j}+1\rbrace}+\sqrt{m_{j}}~\delta_{\mathbf{n},\lbrace m_{j}-1\rbrace}
\right)\right]\left[\sum_{\mathbf{k}}P_{\mathbf{k}\alpha\beta}\right].\notag
\end{eqnarray}